\documentstyle[twocolumn,epsfig,prb,aps,amsmath,amssymb]{revtex}
\begin{document}
\twocolumn[\hsize\textwidth\columnwidth\hsize\csname
@twocolumnfalse\endcsname
\title{Investigation of soliton bound states in the Raman spectrum
of pure and doped spin-Peierls chains}
\author{D. Augier$^a$, E. S\o rensen$^a$, 
J. Riera$^{ab}$ and D. Poilblanc$^a$}
\address{$^a$ Laboratoire de Physique Quantique \& Unit\'e
Mixte de Recherche 5626, Universit\'e P.~Sabatier, 31062 Toulouse, France}
\address{$^b$ Instituto de F\'{\i}sica Rosario, Consejo Nacional de 
Investigaciones 
Cient\'{\i}ficas y T\'ecnicas y Departamento de F\'{\i}sica,
Universidad Nacional de Rosario, Avenida Pellegrini 250, 2000-Rosario,
Argentina}
\date{\today}
\maketitle
\begin{abstract}
We investigate the occurrence of singlet bound states in the Raman spectrum
of dimerized spin 1/2 chains by Exact Diagonalization and
Density Matrix Renormalization Group techniques.
We predict that
several bulk $\delta$-peaks could be observed in pure systems.
Furthermore, we show that new
low energy lines arise from non-magnetic impurity 
doping. These features are interpreted in terms of soliton-antisoliton
and soliton-impurity bound states respectively. 
Energies and spectral weights associated with these bound states
are sensitive to lattice relaxation effects.
Our results are discussed in the context of the inorganic
spin-Peierls compound CuGeO$_3$ and quantitatively compared to
recent Raman experiments.
\end{abstract}
\vskip2pc]

\section{Introduction}
\label{intro}
One-dimensional quantum spin chains have drawn great attention
since the discovery of a spin-Peierls transition in
the inorganic CuGeO$_3$ compound \cite{hase}.
The transition is experimentally inferred from an
isotropic drop of the magnetic susceptibility below a
critical temperature signaling the 
opening of a singlet-triplet
spin gap $\Delta^{01}$ together with a lattice dimerization.

The occurrence of well-defined
magnetic bound states below the excitation continuum 
has been intensively investigated both theoretically and experimentally.
A triplet bound state with $\Delta^{01} \simeq 
16.8$~cm$^{-1}$ has been observed by 
inelastic neutron scattering experiments \cite{arai,ain}. Raman
scattering experiments have identified a sharp singlet resonance
at an energy $\Delta^{00} \simeq 29.7$~cm$^{-1}$ above the ground state
\cite{kuroe,vanloo,muthu,elsprl}.
Recently, Raman experiments on Zn doped CuGeO$_3$
have detected a new singlet bound state at 
$\Delta^{\mathrm{imp}}_1 \simeq 15.0$~cm$^{-1}$ which is close
to the value of the spin gap $\Delta^{01}$
(Ref. \onlinecite{els}). 
Although it is commonly believed that these features signal the
presence of bound states, there is still some controversy about their
interpretation.
It has been proposed that the singlet bound state is a 
magnon-magnon bound state \cite{elsprl}. 
An alternative description of {\it both} the singlet and triplet 
bound states, in
terms of soliton-antisoliton ($s\bar s$) bound states, is also
possible~\cite{affleck,sorensen}. However, the question remains
whether such
$s\bar s$ bound states are observable in Raman spectroscopy.
It is the purpose of the present paper to show that this is indeed the case:
states which are
clearly identifiable as $s\bar s$ or impurity-soliton 
bound states yield a significant 
Raman spectral weight at frequencies roughly coincident 
with the experimental results. 
Hence, a complete understanding of the
experimental results is possible 
within the soliton picture both in pure
and doped materials.

Quasi one-dimensional spin-Peierls compounds are widely described 
in the literature as frustrated dimerized Heisenberg 
chains\cite{riera,castilla}:
\begin{eqnarray}
H=\sum_i [J(1+\delta (-1)^i) \vec{S}_i. \vec{S}_{i+1}
+J_2 \vec{S}_i. \vec{S}_{i+2}],\label{dim}
\end{eqnarray}
where the explicit dimerization $\delta$ arises from some
coupling to the three-dimensional lattice treated at
the mean-field level. 
An extension of this model 
allowing for (slow) variation of the order parameter $\delta$
in space will be discussed latter on.
We first summarize here some well known properties of the
dimerized Heisenberg model (in the following we set $J=1$).

{\it Systems with $\delta=0$.}
A spin gap smoothly opens at $J_{2c} \simeq 0.2412$ 
(Ref. \onlinecite{haldane,okamoto,castilla,eggert}) accompanied by a
spontaneous lattice dimerization. The ground state is then twofold
degenerate. The Majumdar-Ghosh (MG) point \cite{mg}
($J_2=0.5$ and $\delta=0$) is believed to correctly describe the physics
when $J_{2c}<J_2\le1/2$.
The elementary excitations are depicted as solitons ($s$)
and antisolitons ($\bar s)$ \cite{shastry},
topological defects between the two ground state patterns.
No low energy  $s\bar s$ bound states occur in
this model \cite{sorensen} and a continuum of magnetic excitations
begins exactly at $\Delta^{00}=\Delta^{01}=2\Delta_{\mathrm{sol}}$
where $\Delta_{\mathrm{sol}}$ is the minimal energy to create a soliton.

\begin{figure}
\begin{center}
\epsfig{file=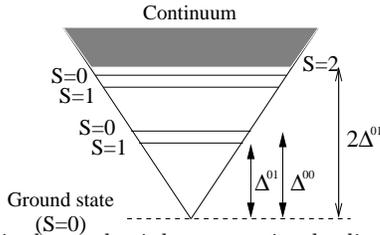,width=5cm}
\caption{Singlet and triplet states in the linear potential.
The continuum begins with a S=2 state (in finite systems).}
\label{potlin1}
\end{center}
\end{figure}

{\it Systems with $\delta\neq 0$.}
For $J_2 > J_{2c}$, a non-zero $\delta$ lifts the degeneracy
between the two ground states and creates a linear 
potential that confines solitons and antisolitons 
into bound states \cite{affleck}. A ladder of $s\bar s$ 
bound states occurs below
the continuum which
begins at $2\Delta^{01}$, {\it i.e.} 
when a higher energy  $s\bar s$ pair  
decays into two pairs \cite{affleck,sorensen} (see Fig. \ref{potlin1}). 
The number of $s\bar s$ bound states
has to be determined numerically and appears to increase as
the dimerization is decreased \cite{affleck,bouz2}.
Turning on a small $\delta$ with $J_2 < J_{2c}$
creates a quite different situation. The explicit dimerization
opens a spin gap $\Delta^{01} \propto \delta^{2/3}$  
transforming the massless spinons of the Heisenberg chains into
massive solitons.
A mapping to the massive Thirring model 
gives evidences for the existence of exactly one triplet and
one singlet bound states with a universal ratio $\Delta^{00}
/\Delta^{01}=\sqrt{3}$ (Ref. \onlinecite{haldane,uhrigschulz,affleck}). 

In this paper, the Raman spectrum of dimerized Heisenberg chains
is numerically investigated by Exact Diagonalization
(ED) and Density Matrix Renormalization Group (DMRG)
calculations.
In section \ref{pure},
we discuss the possibility of observing $s\bar s$ singlet bound states
using Raman spectroscopy on pure systems.
In section \ref{doped}, we consider the effects of 
non-magnetic impurity doping on the Raman spectrum.
Lattice relaxation is studied in section \ref{relaxation}.
Results of sections \ref{pure}, \ref{doped} and 
\ref{relaxation} are obtained for 
parameters close to the MG point, useful to gain insight for a qualitative
understanding.
Finally, the relevance of these results to
CuGeO$_3$ are quantitatively discussed
in section \ref{cuge} in relation with recent experimental results.
 
\section{Pure compound}
\label{pure}

Within the Loudon-Fleury theory \cite{loudon}, the Raman operator 
is written as \cite{muthu,bouz1}:
\begin{eqnarray}
H_R=\sum_i [\vec{S}_i. \vec{S}_{i+1}
+\gamma \vec{S}_i. \vec{S}_{i+2}].\label{ramanop}
\end{eqnarray}
Here, $\gamma$ 
is a microscopic parameter, and terms proportional to the Hamiltonian
have been omitted.
The Raman intensity $I_R$  can then be obtained:
\begin{eqnarray}
I_R(\omega)=\sum_n |\langle \Psi_0|H_R|\Psi_n\rangle|^2
\delta_{\varepsilon}(\omega-E_n+E_0),\label{inte}
\end{eqnarray}
where $|\Psi_n\rangle$ is a complete set of eigenstates with energy $E_n$ 
($n=0$ corresponds to the ground state) and $\delta_{\varepsilon}$
is the lorentzian of width $\varepsilon$.
The Raman operator $H_R$ depends on the
value of $\gamma$. However, 
varying $\gamma$ only makes minute changes for the quantities
of interest {\it at least for the parameter values
used in this paper}. Hence, we shall use the operator
$H_R=\sum_i \vec{S}_i. \vec{S}_{i+1}$ in the following.

\begin{figure}
\begin{center}
\epsfig{file=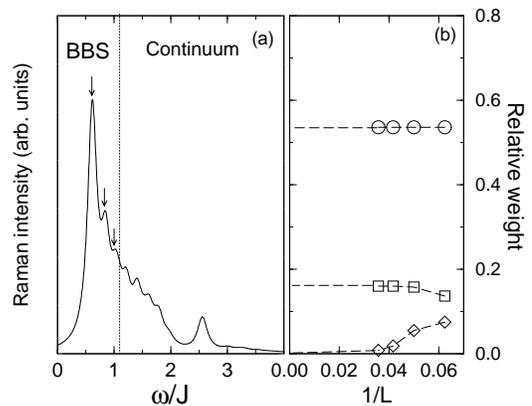,width=7cm}
\caption{(a) Raman intensity calculated for a L=28 site ring with
$J_2=0.5$, $\delta=0.05$ and a broadening $\varepsilon=0.1 J$.
The positions of the three singlet bulk bound state (BBS) from
Ref. \protect\onlinecite{sorensen} are
indicated by arrows. The dashed line shows the onset of the continuum.
(b) Relative weight of the three lowest energy peaks in the Raman spectrum
for $J_2=0.5$ and $\delta=0.1$ as a function of the inverse lattice size.
The presence of two BBS ($\circ,\square$) below
the continuum ($\lozenge$) can then be inferred.}
\label{ramaneven}
\end{center}
\end{figure}

A typical Raman spectrum obtained by ED on a closed
ring of 28 sites
is shown in Fig. \ref{ramaneven}a for $J_2=0.5$,
$\delta=0.05$. 
These 
parameters are chosen for two reasons:
{\it i)} to minimize finite-size effects,
{\it ii)} to study the spectral weight of the two (possibly three)
$s\bar s$ bound states
that numerically have been 
identified for these parameters \cite{sorensen}.
As can be seen in Fig. \ref{ramaneven}a, the Raman intensity is
zero up to $\Delta^{00}$, where the first bulk
bound state (BBS) occurs. This one is by far the most
pronounced feature in the spectrum. The second bound state also has
a significant weight while
the third one is very weak and difficult to distinguish
from the continuum without a precise analysis of finite-size effects.

The relative weights of the peaks are plotted in Fig. \ref{ramaneven}b
for lattice sizes from 16 to 28 sites. Note that the slightly larger
dimerization $\delta=0.1$ used here reduces
finite size effects but  the number of bound
states is lowered from three to two.
At large lattice sizes, the weight of the two 
first peaks is constant and clearly non-zero signalling
their bound state nature in contrast 
to the third peak which vanishes in the thermodynamic limit
indicating that it is part of the continuum.
An equivalent analysis for parameters closer to experimental
values ($J_2\sim 0.36,\ \delta\sim0.014$) (Ref. \onlinecite{riera})
is difficult due to large finite size effects. 
However, the physics should
be similar and it therefore appears reasonable to interpret
the experimentally observed sharp peaks in pure systems
as singlet $s\bar s$ bound states.

\section{Doped compound}
\label{doped}

We now study the effect of introducing in-chain
non-magnetic impurities.
CuGeO$_3$ can be doped with low concentrations of Zn.
The spin 0 Zn substitutes the magnetic Cu$^{2+}$
and in a first approximation the effect of such an impurity
is simply to break the linear chains.
Hence, we can model the effect of Zn doping by studying
chains of different lengths with open boundary conditions.

First we qualitatively analyse the case of odd length chains.
One end of the chain begins with a weak link
$J(1-\delta)$. The chain end will now create a linear potential
capable of binding a soliton.
The ground state is made of a soliton 
bound to the chain end by this linear
potential \cite{laukamp,uhrig,sorensen,els}. 
Other higher lying bound states in the linear potential
may also been identified and
we shall refer to these states as edge bound states (EBS).
These EBS exist up to an energy $\Delta^{01}$ above the ground state,
where it becomes energetically more favorable to create a free
$s \bar s$ pair. As the invariance translation is broken,
a pseudo-continuum due to the $s\bar s$ dispersion begins there.
The lowest lying state of the pseudo-continuum
should consequently be made of the solitonic GS plus
a $s \bar s$ pair and have a spin 3/2 (the impurity bound soliton
and the $s\bar s$ pair do not interact in
the thermodynamic limit).
A two-particle continuum can be observed when two $s \bar s$ pairs
can be created, {\it i.e.} for energies larger than 
$2\Delta^{01}$. 
\begin{figure}
\begin{center}
\epsfig{file=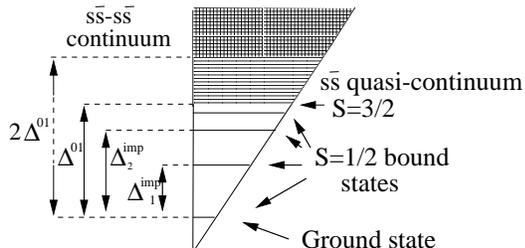,width=7cm}
\caption{The potential well created by an open boundary. The four
soliton-impurity bound states shown here 
correspond to the parameters of
Fig. \ref{spectre}-\ref{etats}-\ref{ramanodd}
($J_2=0.5$, $\delta=0.05$).}
\label{potlin}
\end{center}
\end{figure}

\begin{figure}
\begin{center}
\epsfig{file=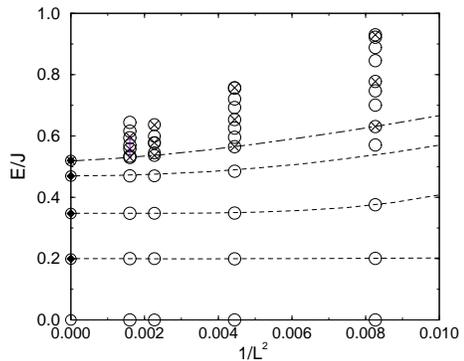,width=6cm}
\caption{Lowest lying $S^z=\frac{1}{2}$ ($\circ$) and $S^z=\frac{3}{2}$
(X) excitations for the $J_2=0.5$ and $\delta=0.05$ Heisenberg chain
as a function of the square inverse of the chain length L 
(up to L=25) obtained by ED. The energy reference is the
GS for each chain length. 
Dashed lines represent
a $(\frac{1}{L^2},\frac{1}{L^3})$ fit and the extrapolations
to infinite sizes are indicated. DMRG extrapolations
are also plotted ($\blacklozenge$). }
\label{spectre}
\end{center}
\end{figure}

We now turn to a more quantitative study.
The number of 
EBS depends sensitively on the parameters of the Hamiltonian
and previous parameters such as $J_2=0.5$ and $\delta=0.05$ should 
be optimal to observe EBS, with small finite size effects.
Fig. \ref{spectre} shows infinite size extrapolations 
of the energies obtained with ED or a DMRG treatment for 
these parameter values. 
Four soliton-impurity bound states 
below the continuum can be clearly 
identified. As discussed before,
the energy of the soliton-impurity
bound state has a finite
upper limit and the onset of the pseudo-continuum
is indicated by a spin 3/2 state.
We have checked that its energy with respect to the GS 
is, in the thermodynamic limit, 
the lowest energy of a $s\bar{s}$ bound state, {\it i.e.} the spin gap 
$\Delta^{01}=(0.5200\pm 10^{-4})J$. 
More explicitly, one finds (see Fig. \ref{potlin}
for the notation)
$\Delta^{\mathrm{imp}}_1\simeq0.20J$, 
$\Delta^{\mathrm{imp}}_2\simeq0.35J$ and 
$\Delta^{\mathrm{imp}}_3\simeq0.47J$.

\begin{figure}
\begin{center}
\epsfig{file=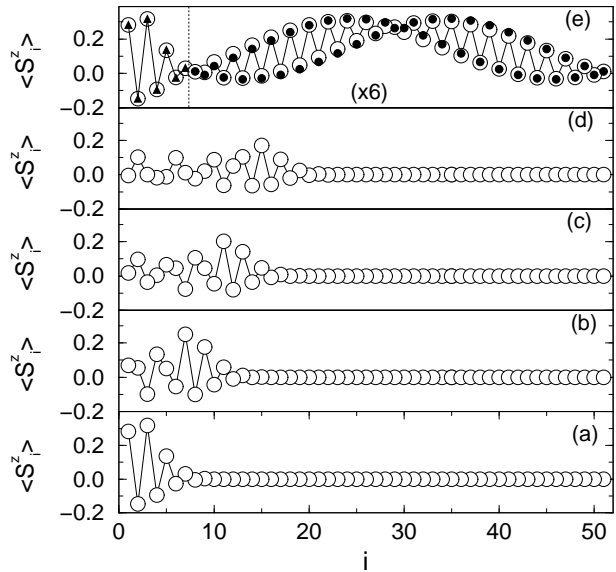,width=8cm}
\caption{$\langle S^z_i \rangle$ as a function
of the site i for the for lowest spin-$\frac{1}{2}$ (a,b,c,d)
and the lowest spin-$\frac{3}{2}$ state (e) for
$J_2=0.5$, $\delta=0.05$ on a 51-site chain using a DMRG algorithm
(m=128).
In (e), a magnification by a factor 6 has been realized 
right to the dashed line. The 7 first sites
of the lowest bound state (a) ($\blacktriangle$) and a 
soliton-antisoliton pair on a 44 site lattice ($\bullet$) are also shown for
comparison.}
\label{etats}
\end{center}
\end{figure}

The wave functions $\langle S^z_i \rangle$ obtained in a DMRG
calculation shown in
Fig.~\ref{etats} clearly support the 
previous interpretation. The four lowest states
(a,b,c,d) show a soliton bound to the weak bond edge, and 
the soliton moves further away from the impurity
when its energy increases. In contrast, in the lowest spin-$\frac{3}{2}$ 
state (Fig.~\ref{etats}e), a $s\bar{s}$ bound pair can be 
clearly identified in addition to the
solitonic GS seen in Fig.~\ref{etats}(a). 

\begin{figure}
\begin{center} 
\epsfig{file=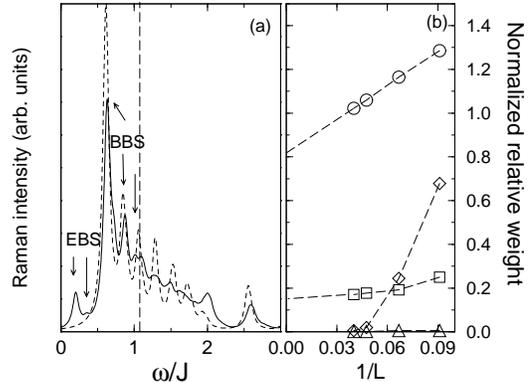,width=7cm}
\caption{(a) Raman intensity calculated for a L=25 site open 
(plain line) and a L=24 site closed chain (dashed line) with
$J_2=0.5$, $\delta=0.05$ and a broadening $\varepsilon=0.05 J$.
The positions of the first two singlet edge bound state (EBS) 
$\Delta^{\mathrm{imp}}_1$, $\Delta^{\mathrm{imp}}_2$ 
are indicated by arrows, together 
with the three BBS of Fig. \ref{ramaneven}. 
The dashed line shows the onset of the bulk continuum.
(b) Relative weight 
of the four lowest energy peaks in the previous Raman spectrum
normalized to the impurity concentration
as a function of $1/L$.
The presence of two EBS ($\circ,\square$) can then be inferred. 
A third one
($\triangle$) has an almost zero although non-vanishing weight.
The fourth peak ($\diamond$) corresponds to the first S=1/2 state
of the quasi-continuum.}
\label{ramanodd}
\end{center}
\end{figure}

As $H_R$ is a spin 0 operator,
transitions between the S=1/2 ground state and higher lying S=1/2
bound states can occur and should be identifiable in the Raman spectrum as 
well-defined peaks at energies $\Delta^{\mathrm{imp}}_i$.
The Raman spectrum of a 25 site chain with open boundary conditions
is shown in Fig.~\ref{ramanodd}a
(plain line) together with the one obtained
for a pure 24 site system (dashed line). The low energy structure
is clearly due to the presence of the impurities and
are signatures of EBS.
The high energy part is the bulk contribution and is
very similar to the one of the pure system (see section \ref{pure}). 
The main effect of introducing impurities is then
to create additional singlet lines
below the singlet-singlet gap $\Delta^{00}$ in the Raman spectrum
as has been experimentally observed \cite{els}. 
The first of the peaks
has an energy $\Delta^{\mathrm{imp}}_1$ 
which we interpret as the energy difference
between the ground state and the first excited bound state of an odd length
chain.

The contribution of the EBS to the total spectrum
as compared to the bulk contribution should scale like the impurity
concentration {\it i.e.} $1/L$.
Hence, in order to correctly identify these states in the relative
spectrum the weights are normalized to the impurity concentration.
The normalized relative weights of the edge bound states
are shown
in Fig. \ref{ramanodd}b for $L$ varying from 11 to 25 sites
and $J_2=0.5$, $\delta=0.05$. In this case it is known that four
edge bound states exist.
Hence, three transition lines should be observable in the Raman
spectrum.
As shown in Fig. \ref{ramanodd}b
the first two transitions between edge bound states have a non-zero weight. 
The third transition
has a very small weight and is almost undetectable.

The presence of the impurity also leads to changes in
the BBS resonances. Since the translation invariance is broken, both
the S=0 and S=1 $s \bar s$ dispersions are reflected at momenta
$q_n=n\pi/L$. Consequently, the BBS peaks are split on an energy range
corresponding to the respective
$s \bar s$ bandwidths although they keep almost all their
spectral weight. 
New small satellites should then be observed in Raman experiments
forming a quasi-continuum.
The possibility of flipping the spin
of the impurity bound soliton may lead in rather small
systems to a peak at the onset of the S=1 $s \bar s$ quasi-continuum. 
The weight of this process
becomes rapidly negligible for $L>15$~sites (see the weight of the
fourth peak in Fig.~\ref{ramanodd}b). 

For even length open chains, there are two possible scenarios.
If the chain begins and ends with a weak bond, the system is
equivalent to the addition of two odd chains. Consequently
there is a soliton-impurity bound state at each chain end and its
edge excitations can be easily inferred from 
the odd system ones.
If both the chain edge bonds are strong no edge bound states exists
and the Raman spectrum should be similar to the bulk spectrum for
pure systems. Note that this latter case with strong bonds at the edges
is favored when lattice relaxation is allowed \cite{hansen}. 

As a consequence, in agreement with what was proposed in 
Ref. \onlinecite{martins}, the previous analysis indicates the presence of
low energy peaks 
below the spin gap 
in inelastic neutron scattering spectra,
due to soliton-impurity bound states.

\section{Role of the lattice relaxation}
\label{relaxation}

So far, the underlying lattice was assumed to be frozen
in a dimerized configuration. However, recent work suggest that
lattice relaxation can be important and lead to qualitatively
new physics. Such effects are particularly crucial 
when inhomogeneities are introduced in the system by, 
for example,
applying a magnetic field~\cite{soliton_lattice,soliton_lattice2} 
or when doping 
with impurities~\cite{hansen}.

In order to go beyond the previous treatment,
we shall assume here that the lattice distortion can 
adjust locally to the magnetic
modulation of the soliton wavefunction.
It has been shown\cite{hh,hansen} that, in this case,
the soliton acquires both a spin and a lattice component.
The description of this magneto-elastic soliton 
can be realized in term of a simple Hamiltonian,
\begin{eqnarray}
H&=&\sum_i [J(1+\delta_i) \vec{S}_i. \vec{S}_{i+1}
+J_2 \vec{S}_i. \vec{S}_{i+2}] + \frac{1}{2}K\sum_i \delta_i^2\nonumber\\
&+& {\tilde K_\perp} \sum_i (-1)^i \delta_i \ ,\label{relax}
\end{eqnarray}
including an elastic constant $K$ and
where the local distortions $\delta_i$ are determined self-consistently
to minimize the total energy (using e.g. an ED method).
The effective interchain coupling $\tilde K_\perp$ introduced here 
is, in fact, related to the physical elastic constant $K_\perp$ 
by the self-consistent relation $\tilde K_\perp=K_\perp\delta$
where $\delta=|\langle\delta_i\rangle_B|$ is the dimerization order
parameter in the bulk (pure system).
Note that the lattice relaxation which we consider here is purely static 
(the distortion $\delta_i$ varies only in space). 
A study of quantum lattice fluctuations is beyond the scope
of this work (for more details on this issue see e.g. 
Ref.~\onlinecite{augier2}).
We then implicitly assume that the time scale of the spin fluctuations
is significantly shorter than the one of the 
lattice (adiabatic approximation).
In this case, the ground state 
$| \Psi_0 \rangle $ contains a spin part
and a distortion $\{ \delta_i^0\}$. Since the Raman operator acts
on the spin part only and the
lattice relaxation
time is much larger than the spin relaxation time scale, 
the eigenstates $| \Psi_n \rangle $ will have the
same distortion $\{ \delta_i^0 \}$. Then, once  $| \Psi_0 \rangle $ 
is obtained with the self-consistent method as explained above, 
the distortion is ``frozen'' for the computation of the spectra 
$I_R(\omega)$ using Eq.~(\ref{inte}).

\begin{figure}
\begin{center}
\epsfig{file=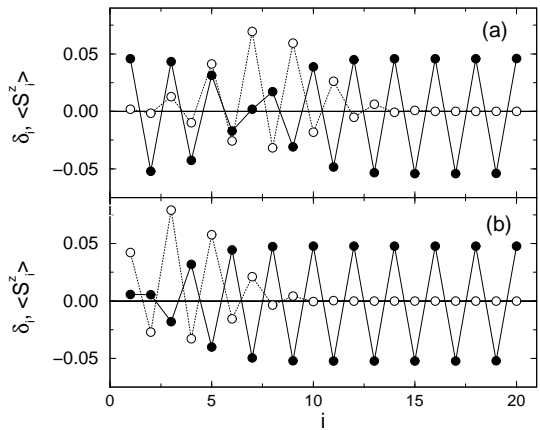,width=7cm}
\caption{Results obtained on a L=21 site chain for the
spin polarization $\langle S_i^z\rangle$ ($\circ$)
and the lattice distortion 
$\delta_i$ ($\bullet$) as a function of the site i (the impurity 
is located on site 0) for $J_2=0.5$. (a) $K=7.7$, $K_\perp \simeq 0$;
(b) $K=11.7$, ${\tilde K_\perp}=0.02$ ($K_\perp\simeq 4$).}
\label{soliton_relax}
\end{center}
\end{figure}

The interchain coupling $K_\perp$ is crucial
in order to create a soliton-impurity
confinement \cite{hansen}. 
The origin of this confinement is clear; in the absence of 
$K_\perp$ the lattice exhibits a solitonic distortion pattern and the
point where $\delta_i$ vanishes coincides with the maximum of the
spin wavefunction $\langle S_i^z \rangle$ as seen 
in Fig.~\ref{soliton_relax}a. Hence, in the region between the impurity and
the soliton the lattice dimerization in the chain is then 
out of phase with the order parameter of the bulk, 
$\langle\delta_i\rangle_B = (-1)^i\delta$.
A finite value of $K_\perp$ then leads to an elastic energy cost proportional
to the impurity-soliton separation. The new equilibrium position is then
obtained when the elastic ``force'' equilibrates the magnetic pressure
that tends to delocalize the soliton away from the wall
as shown in Fig.~\ref{soliton_relax}b.
When ${\tilde K}_\perp\rightarrow\infty$ while the lattice constant $K$ is
tuned to maintain a fixed value of the bulk dimerization $\delta$ 
($K\sim {\tilde K_\perp}/\delta\equiv K_\perp$),
the topological {\it lattice} defect 
is pushed completely towards the impurity end.
The lattice becomes then perfectly dimerized and one recovers the 
previous model. 

\begin{figure}
\begin{center}
\epsfig{file=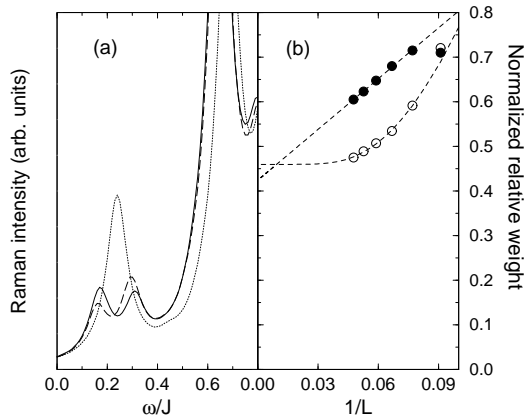,width=7cm}
\caption{(a) Raman intensity calculated on a L=21 open chain
with an adiabatic lattice for
$J_2=0.5$ and a broadening $\varepsilon=0.05 J$. Only the low energy region 
below the BBS $s \bar s$ excitation is shown. 
The dotted, dashed 
and full lines correspond to ${\tilde K}_\perp=0$, $0.1$ and $0.2$
respectively. The values of $K$ ($=7.7$, $9.7$ and $11.7$
respectively) have been chosen to fix the bulk value
$|\langle\delta_i\rangle_B|=\delta \simeq 0.05$.
(b) Relative weight 
of the two lowest energy peaks in one of the previous Raman spectrum
(${\tilde K}_\perp=0.2$, $K=11.7$) normalized to the impurity concentration
plotted vs $1/L$. The dashed lines correspond to fits of the data
using the form $a+b\frac{\exp{(-L/\xi)}}{L}$.}
\label{spectrum_relax}
\end{center}
\end{figure}

We shall focus here on the modification of the Raman spectrum when
$K_\perp$ becomes finite {\it i.e.} when the lattice can relax as described
above. As shown in Figs.~\ref{spectrum_relax}, the low energy
EBS are robust features in the Raman spectrum. 
Note that we have fixed the bulk value of 
$|\langle\delta_i\rangle_B|=\delta\simeq 0.05$ by varying the
parameters $K$ and $K_\perp$ {\it simultaneously} so that a direct
comparison with Fig.~\ref{ramanodd} can be made.
The finite size scaling of the weights in 
Fig.~\ref{spectrum_relax}b unambiguously establishes
the existence of well-defined $\delta$-peaks which can be
interpreted as soliton bound states.

For decreasing mean-field coupling ${\tilde K}_\perp$ important 
changes in the spectrum take place with respect to
the uniformly dimerized chain; (i) the characteristic 
energies of the EBS resonances 
decrease {} 
and (ii) a transfer of spectral weight occurs from the
lowest energy pole to the next one. 
The three spectra shown in Figs.~\ref{spectrum_relax}a 
correspond to ${\tilde K}_\perp=0$, $0.1$ and $0.2$ 
and to (realistic) anisotropy ratios 
$K_\perp/K=0$, $0.21$ and $0.34$ respectively.
The energy shift of the two low-energy peaks to
lower energy is clearly visible (together with the 
transfer of weight) when the lattice becomes more and more
anisotropic. Indeed, for ${\tilde K}_\perp$ decreasing from
$\infty$ (``rigidly'' dimerized chain) to $0$ (isolated 1D chain),
at fixed $\delta=0.05$, the energy 
of the first resonance decreases from $0.20 J$ to $0.13 J$ 
while the second peak moves from $0.35 J$ to $0.24J$. 
As the interchain coupling is varied, a large transfer of weight is
observed in the spectrum. Indeed, 
almost all the low energy spectral 
weight is carried by the first resonance for a large 
${\tilde K}_\perp$
(with a relative weight in the first peak
0.040 vs. 0.0067 in the second one) 
whereas this weight is mainly located in the second peak
for a vanishing interchain coupling (with a relative weight in the first
peak 0.00065 vs. 0.071 in the second one). In this latter case,
only one peak, the second one, can be distinguished in the spectrum 
(see the dotted line
in Fig. \ref{spectrum_relax}a).

\section{Relevance to CuGeO$_3$}
\label{cuge}

\begin{figure}
\begin{center}
\epsfig{file=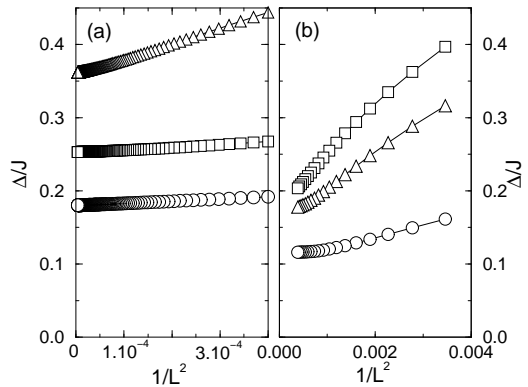,width=7cm}
\caption{Variation of (a) the singlet-triplet $\Delta^{01}$ ($\circ$), 
singlet-singlet $\Delta^{00}$ ($\square$)
and singlet-quintuplet gap $2\Delta^{01}$ 
($\triangle$); (b) the gap between the
ground state and the first $\Delta^{\mathrm{imp}}_1$
($\circ$), second  $\Delta^{\mathrm{imp}}_2$ ($\triangle$) and third 
$\Delta^{01}$ ($\square$) excited state for $J_2=0.36$ and
$\delta=0.014$ as a function of the inverse square chain length
using a DMRG algorithm for open chains.}
\label{scaling}
\end{center}
\end{figure}

So far all calculations have been performed for parameters
close to the MG point in order to minimize finite size effects
and clearly identify the bound states.
We now discuss the implications of our results for more
realistic parameter values relevant to CuGeO$_3$.
Parameters such as $J\simeq 160$~K,
$J_2 \simeq 0.36$ and $\delta \simeq 0.014$ have been inferred
by ED from high temperature susceptibility measurements and 
the zero temperature spin gap value \cite{riera,bouz1,low}.
Raman spectra obtained by ED would have very large
size effects for these parameters, making an
interpretation very difficult. Indeed, previous studies
for $J_2=0$ or $J_2=0.24$ have seen no indication of
soliton-antisoliton bound state \cite{muthu,navet}.
Hence, we turn to DMRG techniques,
first omitting possible lattice relaxation effects.
Although the associated spectral
weights cannot be obtained, the different
gaps can be precisely evaluated using
even open chains:
$\Delta^{01}/J \simeq 0.1802$,  $\Delta^{00}/J \simeq 
0.2529$   and $\Delta^{02}/J \simeq 0.3606$ (see Fig. \ref{scaling}a). 

\begin{figure}
\begin{center}
\epsfig{file=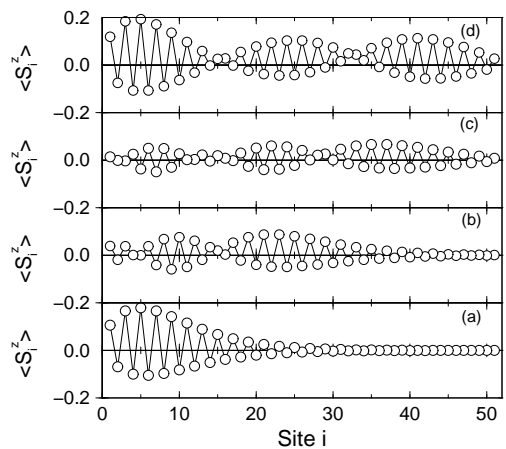,width=6.5cm}
\caption{DMRG results for
$\langle S_i^z \rangle$ as a function of the site i for the three
lowest S=1/2 (impurity-soliton bound states: a, b, c) and the
lowest S=3/2 state (onset of the 
quasi-continuum: d) on a 51 site chain beginning
with a weak link for $J_2=0.36$, $\delta=0.014$.}
\label{wave}
\end{center}
\end{figure}

Using the same parameter values, it appears that
three impurity-soliton bound states
(Fig. \ref{wave}a,b,c) occur for odd open chains.
These three bound states have a very extended but still
localized wavefunction,
signalled by an exponentially decaying part of $\langle S^z_i \rangle$
far from the chain end.
For instance, the
exponential part of the ground state (Fig. \ref{wave}a)
is $\sim \exp(-r/6.8)$.
In contrast, the first S=3/2 state shown in Fig. \ref{wave}d 
is made of a soliton close to the impurity (identical to the
ground state (a) in the thermodynamic limit) 
plus a soliton-antisoliton pair which extends all over the rest of the
chain. Note that the last bound state appears immediately
below the first S=3/2 state. This is clear for short and intermediate
lengths, however, this state could possibly become part of the 
$s \bar s$ quasi-continuum
when $L\to\infty$. Our data seem to indicate that this is not the case. 
The two gaps associated with
these three impurity-soliton bound states have also
been evaluated by DMRG (Fig. \ref{scaling}b):
$\Delta^{\mathrm{imp}}_1/J \simeq 0.116$, and 
$\Delta^{\mathrm{imp}}_2/J \simeq 0.175$ (just below the quasi-continuum).
The energy difference between the ground state
and the third excited state, which has a spin 3/2
and is the first state of the quasi-continuum,
gives approximately the spin gap $\Delta^{01}$
(see Fig. \ref{scaling}b), as we have explicitly checked.
Equivalently,
using $J\simeq 160$~K$\simeq 112$~cm$^{-1}$, one obtains:
$\Delta^{01}\simeq 20$~cm$^{-1}$, $\Delta^{00} \simeq 28$~cm$^{-1}$,
of the same order as the experimental values $\simeq 17$~cm$^{-1}$
and $30$~cm$^{-1}$ respectively. The small deviation
between theory and experiments may be attributed to other
effects not included in this paper such as 
magnetic interchain coupling. For the doped systems our results are
$\Delta^{\mathrm{imp}}_1\simeq 13~$cm$^{-1}$, in surprisingly
good agreement with the experimental result $15$~cm$^{-1}$ (Ref. 
\onlinecite{els}). The ratio $\Delta^{\mathrm{imp}}_1/
\Delta^{01}$ is about $0.6$, which is very similar to what has been
found using a particle in a linear potential \cite{els}.
However, according to our analysis, another EBS should be
observed at $\Delta^{\mathrm{imp}}_2
\approx 20$~cm$^{-1}$, contrary to the linear
potential approach that predicts only
one \cite{els}. 
However, the spectral weight of this state 
might be very small.

For an anisotropic lattice (see model~(\ref{relax})),
another alternative explanation might have to be invoked. Indeed,
in that case, significant lattice relaxation would shift the
poles to smaller energies. One is then tempted to attribute
the experimental resonance at $15$~cm$^{-1}$ to the
{\it second} theoretical EBS. In this scenario, another
EBS is expected with a small weight at energies below $10$~cm$^{-1}$.
A direct comparison with the experimental spectrum \cite{els} may
corroborate this prediction but the signal noise prevents 
a definitive conclusion. Furthermore, detection of resonances
in Raman scattering experiments at very low energy is difficult
due to Brillouin scattering.

In conclusion, we have shown that $s\bar s$ bound states should
be easily identifiable in experimental Raman spectra on pure systems.
Consequently, it is reasonable to interpret the experimentally observed 
line at $29.7$~cm$^{-1}$ as a $s\bar s$ bound state.
Doping with chain breaking impurities should give rise to a number
of edge bound states with non-negligible Raman weight. At least one 
such transition has already been experimentally observed~\cite{els}
in CuGeO$_3$.

We thank IDRIS (Orsay) for allocation of CPU time on the C94 and C98
CRAY supercomputers and rs6000 workstations.
We also acknowledge funding from the 
ECOS-SECyT A97E05 programme.
J.R. thanks C.N.R.S. (France) for financial support.

\end{document}